\documentclass{aastex63}
\usepackage[T1]{fontenc}
\def\simgr{\,\hbox{\hbox{$ > $}\kern -0.8em \lower 1.0ex\hbox{$\sim$}}\,}
\def\simle{\,\hbox{\hbox{$ < $}\kern -0.8em \lower 1.0ex\hbox{$\sim$}}\,}

\shortauthors{THORSTENSEN}
\shorttitle{LAMOST CVs}

\begin{document}
\title{Follow-up Studies of Five Cataclysmic Variable Candidates
Discovered by LAMOST
}

\author[0000-0002-4964-4144]{John R. Thorstensen
}

\affil{Department of Physics and Astronomy,
6127 Wilder Laboratory, Dartmouth College,
Hanover, NH 03755-3528}

\begin{abstract}
We report follow-up observations of five cataclysmic variable 
candidates from LAMOST published by \citet{hou20}.  
LAMOST J024048.51+195226.9 is the most unusual of the five;
an early-M type secondary star contributes strongly to 
its spectrum, and its spectral and photometric behavior 
are strikingly reminiscent of the hitherto-unique 
propeller system AE Aqr.  We confirm that a 7.34-hr period
discovered in the Catalina survey data \citep{drake14} is orbital.  
Another object, LAMOST J204305.95+341340.6 appears to be a near 
twin of the novalike variable V795 Her, with an orbital period 
in the so-called 2-3 hour ``gap''.  LAMOST J035913.61+405035.0 
is evidently an eclipsing, weakly-outbursting dwarf nova with 
a 5.48-hr period.  Our spectrum of LAMOST J090150.09+375444.3 is 
dominated by a late-type
secondary and shows weak, narrow Balmer emission moving 
in phase with the absorption lines, but at lower amplitude;
we do not see the \ion{He}{2} $\lambda$4686 emission evident
in the published discovery spectrum.  We again confirm that a
period from the Catalina data, in this case 6.80 hr, is orbital.
LAMOST J033940.98+414805.7 yields a radial-velocity 
period of 3.54 hr, and its spectrum appears to be typical
of novalike variables in this period range.  The spectroscopically-
selected sample from LAMOST evidently includes some interesting
cataclysmic variables that have been unrecognized until 
now, apparently because of the relatively modest range of 
their photometric variations.

\end{abstract}


\section{Introduction}

Cataclysmic variable stars (CVs) are binary systems in which a
white dwarf accretes matter from a companion via Roche
lobe overflow.  The companion usually resembles a main-sequence
star, but often shows significant differences, from 
some combination of nuclear evolution and the effects 
of having transferred significant mass \citep{knigge06,knigge11}.  
CV orbital periods $P_{\rm orb}$ range from $\sim 5$ min for the
most extreme double-degenerate systems, up to several days for
systems that have evolved significantly,
but more typically range from $\sim 80$ min up to $\sim 10$ hr
for hydrogen-burning secondaries.  \citet{warner95} gives
an overview of these objects; \cite{goliasch15} and
\cite{kalomeni16} describe models of CV evolution.

CVs fall into several subclasses, some of them overlapping.
Many are dwarf novae, which from time to time undergo 
brightenings (called {\it outbursts}).  The outbursts 
are thought to be caused by an instability triggered when 
the accretion disk reaches a critical density.  In another
class, called novalike variables, the mass-transfer rate from
the secondary star is large enough to maintain the 
disk in a bright state; some novalikes fade into low
states, apparently when mass transfer slows or stops. 
In some CVs the white dwarf is highly magnetized,
which affects the accretion flow.  In AM Hercuis stars, also
called {\it polars}, no disk forms, 
the white dwarf co-rotates with the 
orbit, and the magnetic field channels the accretion flow
onto the magnetic poles of the white dwarf. 
If the white dwarf field is somewhat smaller, the accretion
disk is not entirely disrupted, and
the white dwarf does not
co-rotate with the orbit.  In such systems, called 
DQ Herculis stars or {\it intermediate polars}, the brightness
is often modulated at the white dwarf rotation period
or an orbital sideband.  Some intermediate polars
show dwarf nova outbursts (for example,
EX Hya; \citealt{beuermann03}), but more typically they are novalike variables.

CVs are discovered through several different channels, all
of which are biased in one way or another.
Most cataloged CVs are dwarf novae. Dwarf novae are usually
discovered because of their outbursts, which trigger
variability surveys such as the Catalina surveys 
\citep{crts,breedt14}, ASAS-SN \citep{shappee14}, MASTER \citep{lipunov10}, 
and ATLAS \citep{tonry18}. 
Many CVs have been discovered as X-ray sources, 
especially the magnetic subtypes
(see \citealt{schwope18} and references therein).  
Most CVs have unusual colors, and were discovered
in spectroscopic or variability 
follow-ups of color-selected objects \citep{green82,
szkody11}.
CVs can have strong emission lines (though not all
do), which stand out in spectroscopic surveys \citep{aungwerojwit06,
witham08}.
Even with all these means of discovery, the sample of known CVs 
is apparently not representative; as one example, 
\citet{pala20} examine a volume-limited sample
of CVs within 150 pc of earth, as determined by
Gaia DR2 parallaxes, and find that over 30 
per cent are magnetic CVs, far in excess of their
numbers in other catalogs.  It is therefore always
a welcome development when new CV samples become
available using novel selection methods.

Recently, \citet{hou20} presented just such a sample.
They examined spectra from the Large Sky Area Multi-Object 
Fiber Spectroscopic Telescope (LAMOST) using 
machine-learning techniques, and selected 245 CV candidates 
from among the millions spectra produced by the survey.
While most of their sample had already been identified 
as CVs, 58 of their candidate CVs apparently had 
not been previously been recognized as such.

In this paper we present initial optical 
follow-up studies of five of these new objects.
We set out to verify that they are CVs, assess their 
subtypes, determine reliable orbital periods 
$P_{\rm orb}$ when possible.

Table \ref{tab:targets} gives basic information
on the stars we observed; for brevity, we will sometimes
abbreviate their names.  The distances quoted are 
simple inverses of the Gaia parallaxes, which 
have relative errors small enough to be accurate enough 
for our purposes.  We used the Gaia distances together with 
the \citet{green19} 3-D reddening maps\footnote{See
http://argonaut.skymaps.info/} to estimate the tabulated
reddenings $E(g-r)$.  The approximate absolute $G$ magnitudes are 
calculated using the central value of the 
distance, and assuming that (a) reddening in 
$G$ is comparable to reddening in $V$, and (b) 
$E(B-V) \sim  E(g-r)$ (see the discussion
at {\tt http://argonaut.skymaps.info/usage\#versions}).
Note that $M_G$ should be comparable to $M_V$, since
$G$ and $V$ magnitudes differ by only a few tenths for 
normal blue ($B - V < 0.5$) stars (see Fig. 15 of 
\citealt{carrasco16}).

The plan of the paper is as follows.
Section \ref{sec:techniques} describes
our observation, reduction and analysis procedures.  
The results for the individual stars 
are given in Section \ref{sec:stars}, in order of
right ascension.  Table \ref{tab:velocities} 
gives the measured radial velocities, and 
Table \ref{tab:sinefits} gives parameters of sinusoidal
fits to the velocities.  We summarize our results
in Section \ref{sec:discussion}.

\section{Techniques}
\label{sec:techniques}

All the data presented here were taken 
in 2019 December and 2020 January, at 
MDM Observatory on Kitt Peak, Arizona.  

All our spectra are from the 2.4m Hiltner
reflector with the Ohio State Multi-Object 
Spectrometer (OSMOS; \citealt{martini}) used in 
single-slit mode.   We used the ``blue" disperser, 
with the 1.1-arcsecond ``inner'' slit, yielding
$\sim 3.1$ \AA\ resolution (FWHM) from 3980 to 6860 \AA,  
and a dispersion of 0.7 \AA\ per 15 $\mu$m pixel.
The CCD detector we used has four amplifiers, so
we used a python script to compute and subtract
bias levels from each quadrant.   For spectroscopic
flat fields we used spectra of an incandescent bulb.
As a check, we also created flat field images from spectra
of the sky taken soon after sunset.  To eliminate 
the strong solar spectrum from these, we  
divided the data in each column (i.e., at each wavelength)
by a polynomial fitted to that column.  The sky-derived
flats were nearly identical to those from the bulb, but
had slightly poorer signal-to-noise.

To extract one-dimensional
spectra from the images, we used a new python
implementation of the variance-weighted algorithm
described by \citet{horne86}\footnote{The program 
is available at
https://github.com/jrthorstensen/opextract.}.
To calibrate the wavelength scale,
we took exposures of Hg, Ne, and Xe 
comparison lamps in the zenith during the day,
and then bracketed our science exposures with 
short Hg-Ne exposures to track the drift in the 
zero point and the dispersion (which did change
significantly) with telescope position and 
ambient temperature.  The resulting calibration
was typically accurate to $\sim 10$ km s$^{-1}$,  
judging from night-sky features \citep{osterbrock16}.
When the sky was clear and the seeing acceptable,
we observed flux standard stars from which we
derived a flux calibration.

We measured emission-line radial velocities 
by convolving the line profiles with antisymmetric functions 
and looking for the zero of the convolution
\citep{sy80,shafter83}.  With this method one
can vary the width and functional form of the 
antisymmetric function to emphasize or exclude
different parts of the line profile.  Each pixel
in the spectrum has an uncertainty estimate based
on the gain, background, read noise, and flux;
we propagated those through the measurement 
process to estimate the uncertainty $\sigma_i$ in 
each velocity $v(t_i)$, or rather a lower limit, since
the estimate does not include systematic errors.   

The spectra of several of these the objects showed 
contributions from late-type stars.
To measure the velocities of the late-type
component, we cross-correlated the 
spectrum against late-type template spectra, using the IRAF 
task {\tt fxcor}, after masking out regions
around emission lines.  For the templates, we used 
averaged spectra of late-type stars with
accurately-known velocities; the individual
stars had been shifted into the rest frame 
before averaging.  The template spectra were originally
taken with the {\it modspec} spectrograph on 
the 2.4 m telescope.

To search for periods, we constructed a dense grid of 
trial frequencies covering from 
near zero to 20 cycles per day, fit
least-squares sinusoids at each frequency,
and plotted $1/\chi^2$ {\it versus}
frequency, where 
\begin{equation}
\chi^2 = \sum_i {(v(t_i) - c_i)^2 \over \sigma_i^2},
\end{equation}
$v(t_i)$ is the velocity measured at time $t_i$,
and $c_i$ is the expected value at $t_i$ 
computed using the best-fit sinusoid.
Even in well-sampled data these ``residual-grams''
invariably show possible fits at several frequencies
separated by 1 cycle d$^{-1}$, because velocities
from a single site cannot be taken during daytime;
this causes ambiguity, or aliasing, in the daily cycle count.
To mitigate this, we included observations 
with as wide a spread of hour angles -- essentially
times {\it modulo} one day -- as we could manage.
At large hour angles, differential atmospheric refraction 
\citep{filippenko82} can be severe, so when necessary
we rotated the instrument to orient the slit near the parallactic
angle.  

We fit the radial velocity time series with 
least-squares sinusoidal functions
of the form 
\begin{equation}
v(t) = \gamma + K \sin\left[{2 \pi(t - T_0) \over P}\right],
\end{equation}
and estimated parameter uncertainties from the goodness 
of fit.  Table~\ref{tab:sinefits} shows the 
parameters for all the targets.

To acquire a spectroscopic target with OSMOS, one first
obtains a direct image of the field by removing the 
disperser and slit from the beam (configuring the instrument
as a reducing camera) and
taking a brief (typically 20 s) exposure.  Using this image, one 
fine-tunes the telescope position so that after the slit
is returned to the beam, the target will fall in the slit.
To enhance the scientific
value of these acquisition images, we took them with 
a Sloan $g$ filter.  The images covered 
4.6 $\times$ 18.4 arcmin, and therefore 
included many field stars as well as the target.
To efficiently calibrate the targets' $g$ magnitudes,
we developed a procedure that (1) automatically generates
an astrometric solution by matching the stars in the image
to their entries in the Pan-STARRS 1 catalog, 
(2) performs aperture photometry on the field star images,
(3) finds the mean offset between the aperture-photometry 
magnitudes and Pan-STARRS $g_{\rm PSF}$, and (4) 
applies this offset to the target's aperture magnitude.
The scatter in the magnitude offset was small, typically 
only a few hundredths of a magnitude for well-exposed 
stars.  The procedure also yields a measure of the
image quality, (which for some of our observations was 
remarkably poor); the size of the magnitude offset
is also sensitive to seeing and transparency variations.  

For some of our targets, we also
obtained differential time-series photometry
using the 1.3-meter McGraw-Hill
telescope.  In 2019 December we used a SITe 
CCD detector cropped to a $256 \times 256$ sub-array
of $0''.508$ pixels, and in 2020 January we
used an Andor IKON camera with a 
frame-transfer CCD.  All the images were essentially
`white light', being filtered only by a Schott 
GG420 glass filter, which passes $\lambda > 4200$ \AA\  
and therefore helps suppress scattered moonlight.
In each image we measured instrumental magnitudes for the program
star, a comparison star, and several check stars  
using the aperture photometry code in 
the IRAF implementation of DAOPHOT.  While the 
raw data are uncalibrated and differential, we
adjust them here to approximate standard magnitudes using
catalog magnitudes of the comparison stars. 
Because the original images were basically unfiltered, we
estimate the zero point to be uncertain by several
tenths of a magnitude.

\begin{deluxetable}{llrrlrrr}
\label{tab:targets}
\tablewidth{0pt}
\tablecolumns{7}
\tablecaption{List of Objects}
\tablehead{
\colhead{Name} &
\colhead{$\alpha_{\rm ICRS}$} &
\colhead{$\delta_{\rm ICRS}$} &
\colhead{$G$} &
\colhead{$1/\pi_{\rm DR2}$} &
\colhead{$E(g - r)$} &
\colhead{$M_G$} \\
\colhead{} &
\colhead{[h:m:s]} &
\colhead{[d:m:s]} &
\colhead{} &
\colhead{[pc]} &
\colhead{} &
\colhead{} \\
}
\startdata
LAMOST J024048.51+195226.9 & 02:40:48.531 & +19:52:26.96 & 16.8 &  $627 \pm  36$ & 0.02 & 7.8 \\
LAMOST J033940.98+414805.7 & 03:39:40.991 & +41:48:05.69 & 15.2 &  $914 \pm  38$ & 0.17 & 4.9 \\
LAMOST J035913.61+405035.0 & 03:59:13.625 & +40:50:35.08 & 17.4 &  $874 \pm  94$ & 0.34 & 6.7 \\
LAMOST J090150.09+375444.3 & 09:01:50.119 & +37:54:44.22 & 16.9 &  $525 \pm  28$ & 0.01 & 8.3 \\
LAMOST J204305.95+341340.6 & 20:43:05.955 & +34:13:40.73 & 15.3 &  $992 \pm  27$ & 0.35 & 4.3 \\
\enddata
\tablecomments{The celestial
coordinates, mean $G$ magnitudes, and distances are 
from the GAIA Data Release 2 (DR2; \citealt{GaiaPaper1,GaiaPaper2}).  
Positions are referred to the
ICRS (essentially the reference frame for J2000), and the catalog
epoch (for proper motion corrections) is 2015.  The distances and
their error bars are the inverse of the DR2 parallax $\pi_{\rm DR2}$,
and do not include any corrections for possible systematic errors.
See text for discussion of the last columns ($E(g-r)$ and $M_G$).
}
\end{deluxetable}

\begin{deluxetable}{llrrrr}
\label{tab:velocities}
\tablewidth{0pt}
\tablecolumns{6}
\tablecaption{Radial Velocities}
\tablehead{
\colhead{Target} &
\colhead{Time} &
\colhead{$v_{\rm abs}$} &
\colhead{$\sigma(v_{\rm abs})$} &
\colhead{$v_{\rm emn}$} &
\colhead{$\sigma(v_{\rm emn}$)} \\
\colhead{} &
\colhead{} &
\colhead{(km s$^{-1}$)} &
\colhead{(km s$^{-1}$)} &
\colhead{(km s$^{-1}$)} &
\colhead{(km s$^{-1}$)} \\
}
\startdata
LAMOST J024048.51+195226.9   & 58829.8120 &  $  -44$ & $  25$ &  \nodata & \nodata \\ 
LAMOST J024048.51+195226.9   & 58829.8194 &  $   41$ & $  33$ &  \nodata & \nodata \\ 
LAMOST J024048.51+195226.9   & 58829.8286 &  $   16$ & $  39$ &  \nodata & \nodata \\ 
LAMOST J024048.51+195226.9   & 58829.8360 &  $  212$ & $  47$ &  \nodata & \nodata \\ 
LAMOST J024048.51+195226.9   & 58830.8746 &  $  111$ & $  35$ &  \nodata & \nodata \\ 
LAMOST J024048.51+195226.9   & 58830.8834 &  $   38$ & $  20$ &  \nodata & \nodata \\ 
\enddata
\tablecomments{Radial velocities used in this study.  The full table is published as
a machine-readable table; the first few lines are shown here to indicate its form
and content.}
\end{deluxetable}

\begin{deluxetable}{llllrrrc}
\label{tab:sinefits}
\tablewidth{0pt}
\tablecolumns{8}
\tablecaption{Parameters of Best Fit Sinusoids}
\tablehead{
\colhead{Target} &
\colhead{Line type} &
\colhead{$T_0$\tablenotemark{a}} &
\colhead{$P$} &
\colhead{$K$} &
\colhead{$\gamma$} &
\colhead{$N$} &
\colhead{$\sigma$} \\ 
\colhead{} &
\colhead{} &
\colhead{} &
\colhead{(d)} &
\colhead{(km s$^{-1}$)} &
\colhead{(km s$^{-1}$)} &
\colhead{} &
\colhead{(km s$^{-1}$)} \\
}
\startdata
LAMOST J024048.51+195226.9 & abs. & 58836.845(2) & 0.305684\tablenotemark{b}  &  250(8) & $ 33(6)$ & 48 &  30 \\[1.2ex] 
LAMOST J033940.98+414805.7 & emn. & 58832.686(3) & 0.14712(18) & 182(21) & $ 41(15)$ & 58 &  50 \\[1.2ex]
LAMOST J035913.61+405035.0 & abs. & 58832.868(5) & 0.228344\tablenotemark{c}  & 215(28) & $-53(20)$ & 16 &  52 \\
LAMOST J035913.61+405035.0 & emn. & 58833.000(5) & 0.228344\tablenotemark{c}  & 190(33) & $-40(21)$ & 35 &  72 \\[1.2ex]
LAMOST J090150.09+375444.3 & emn. & 58836.1463(17) & 0.283387\tablenotemark{b} &  101(7) & $ 8(3)$ & 20 &  12 \\ 
LAMOST J090150.09+375444.3 & abs. & 58836.1522(13) & 0.283387\tablenotemark{b} &  190(8) & $ 6(5)$ & 20 &  16 \\ [1.2ex]
LAMOST J204305.95+341340.6 & emn. (best) & 58832.6260(18) & 0.10777(10) &  208(19) & $-27(15)$ & 28 &  50 \\
LAMOST J204305.95+341340.6 & emn. (alternate) & 58832.625(2) & 0.12078(14) &  210(20) & $-13(15)$ & 28 &  52 \\
\enddata
\tablecomments{Fits to radial velocities.  The last two columns give the number
of points fitted and the root-mean-square residual. Uncertainties are given in 
parentheses, in units of the last digits quoted.}
\tablenotetext{a}{Barycentric Julian Date, minus 2,400,000., in the UTC time 
scale.}
\tablenotetext{b}{Period fixed at the value given by \citet{drake14}.}
\tablenotetext{c}{Period fixed by the provisional eclipse ephemeris (Eqn.~\ref{eqn:lam0359provisional}).}
\end{deluxetable}

\section{The Individual Stars}
\label{sec:stars}


\subsection{LAMOST J024048.51+195226.9}
\label{subsec:lam0240}

\begin{figure}
\includegraphics[width=7.0 truein]{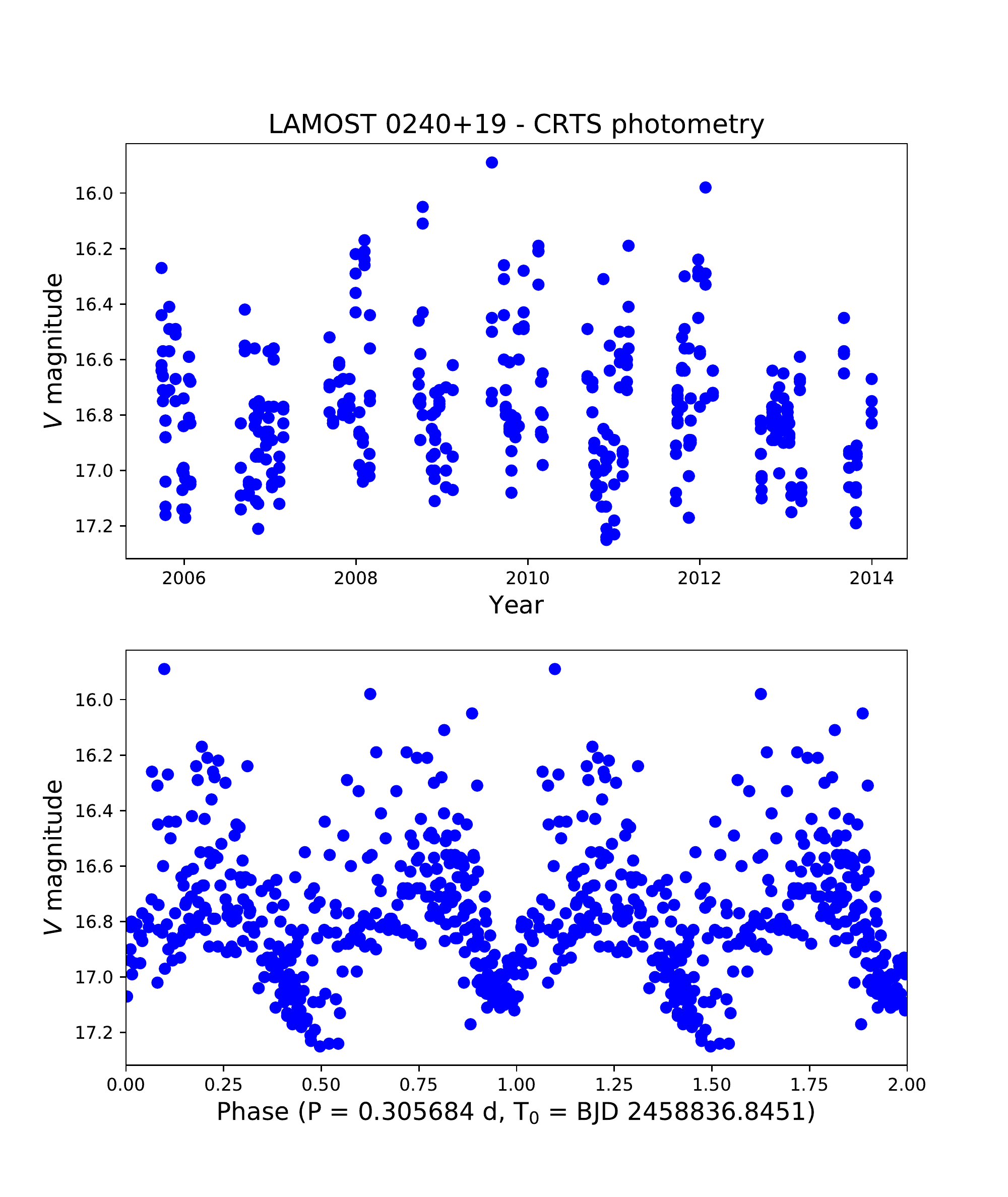}
\caption{  
Archival photometry of LAM 0240+19 from the CRTS DR2. 
The top panel shows the data as a function of time, and 
the bottom panel shows the data folded on the period
found by \citet{drake14}.  Phase zero corresponds to 
inferior conjunction of the secondary star, as deduced
from the absorption-line radial velocities.
}
\label{fig:lam0240catalina}
\end{figure}

The LAMOST spectrum \citep{hou20} shows 
strong Balmer lines, but rather weak features
of \ion{He}{1} and no evident \ion{He}{2} lines;
the continuum shows wide absorption bands indicating
that an M star contributes a significant 
fraction of the light.  The Catalina 
Real Time Survey light curve \citep{crts}, 
shown in Fig.~\ref{fig:lam0240catalina}, shows
the source most often just brighter than 17th
magnitude, but often
brighter than 16.5, and never fainter than about
17.2.  There is no evident pattern in the long-term 
variability, and in particular distinct outbursts 
are not present.  However, \citet{drake14} searched
the Catalina data for periodic variations and 
discovered a periodicity at 0.3056840 d, or
7.33555 hr; the lower panel of Fig.~\ref{fig:lam0240catalina}
shows the same data, folded on this period. 

\begin{figure}
\includegraphics[width=7.8 truein, trim=0.5cm 1cm 0 2cm,clip]{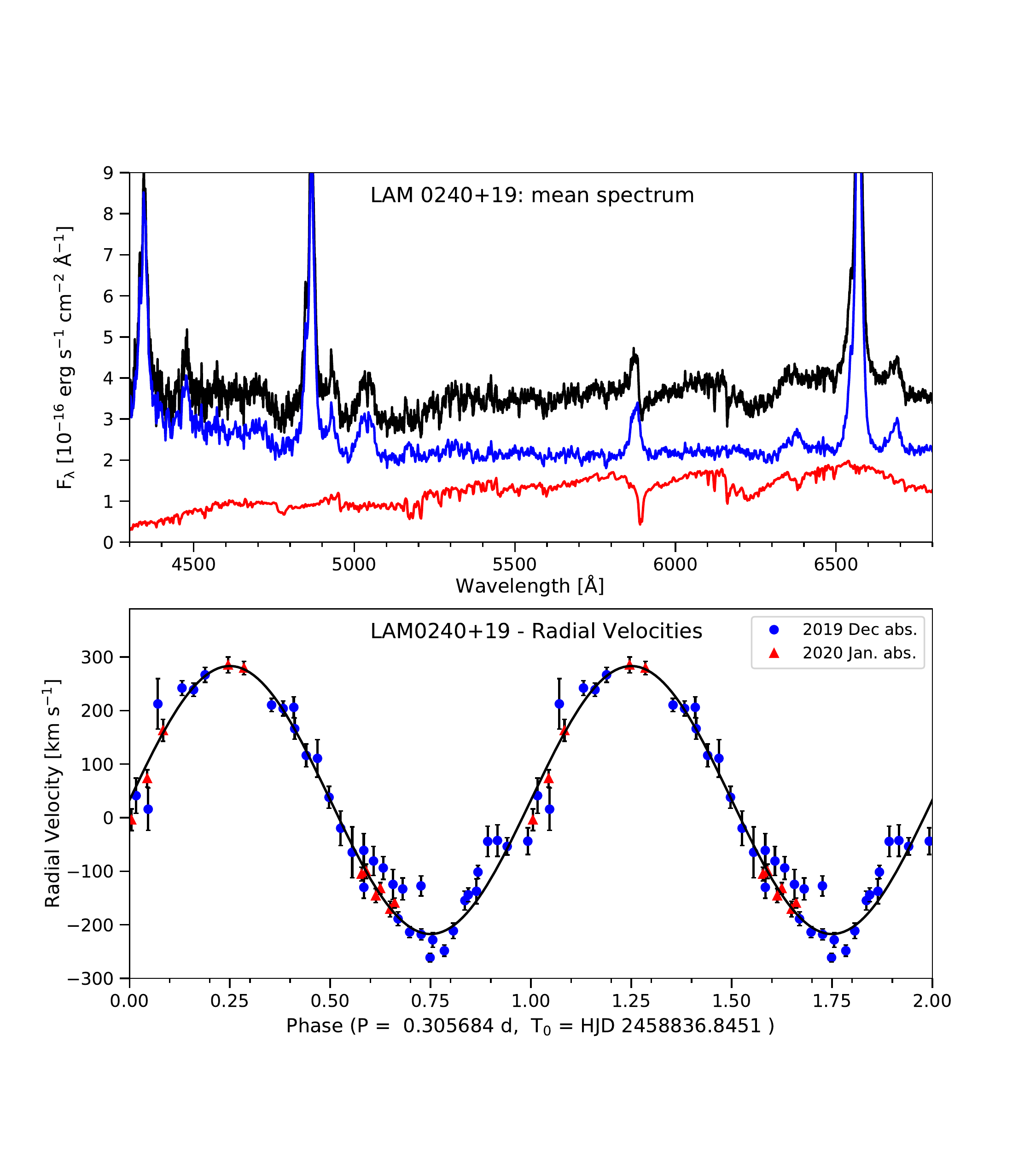}
\caption{  
Upper panel: The mean spectrum of LAM 0240+19. The top
trace (black) is the observed spectrum averaged in the 
rest frame of the late-type star; the middle trace
(blue) shows the same data after subtraction of the
spectrum shown in bottom trace (red), which is a 
scaled spectrum of the M1.5-type dwarf Gliese 15a.
Lower panel:  Radial velocities of the secondary star, folded
on the ephemeris shown.  The best-fitting sinusoid is
plotted, and the data are repeated through a second
cycle for continuity.
}
\label{fig:lam0240montage}
\end{figure}

We observed this star most extensively in 
2019 December.  Our mean spectrum 
(Fig.~\ref{fig:lam0240montage}, top panel) 
resembles the LAMOST spectrum.
Before averaging, 
the individual spectra were shifted to the
rest frame of the late-type star using the 
orbital ephemeris (see below).
To characterize the late-type component, we
subtracted spectra of M-dwarfs classified by
\citet{boeshaar76}, which had been shifted
to their own rest frames using velocities from
\citet{marcy87}, and varied the spectral 
type and flux of the subtracted spectrum to
obtain the best cancellation of the late-type
features.  From this we estimate the secondary
spectral type to be M1.5 $\pm$ 1 subclass.  



We searched for $P_{\rm orb}$ using the
the H$\alpha$ emission line velocities, but these
showed significant scatter and did not yield a unique
period.  In contrast, the cross-correlation velocities 
from the absorption lines varied smoothly 
on a period of 7.34(2) hr.  This was accurate enough
to connect the cycle count to spectra
obtained in 2020 January; the combined data give 
7.3371(8) hr, consistent with the photometric period 
from \citet{drake14} within the mutual uncertainty.  
The \citet{drake14} period is more precise than ours
due to the long time base, so we adopt this as 
the orbital period.  The
lower panel of Fig.~\ref{fig:lam0240montage} shows the 
absorption velocities folded on $P_{\rm orb}$.

The absorption-line velocities trace the motion
of the secondary star, and therefore establish the
binary phase for this non-eclipsing system.
The epoch $T_0$ corresponds to blue-to-red 
crossing of the secondary star's velocities,
which is the phase at which the secondary would
eclipse the accreting object if the binary inclination 
were higher.  The folded Catalina magnitudes
(Fig.~\ref{fig:lam0240catalina}) show a pattern
of two maxima and two unequal minima per orbit, 
evidently so-called ellipsoidal variation caused
by the tidal distortion of the secondary star.  
The deeper of the two minima occurs near phase
$\phi = 0.5$, as expected from the deeper 
gravity darkening on the side facing the 
L1 point.  The numerous brighter points in the 
CRTS light curve suggest short-lived
flares of up to about 1 mag.

\begin{figure}
\includegraphics[width=7.0 truein, trim = 0.5cm 3cm 1cm 3cm, clip ]{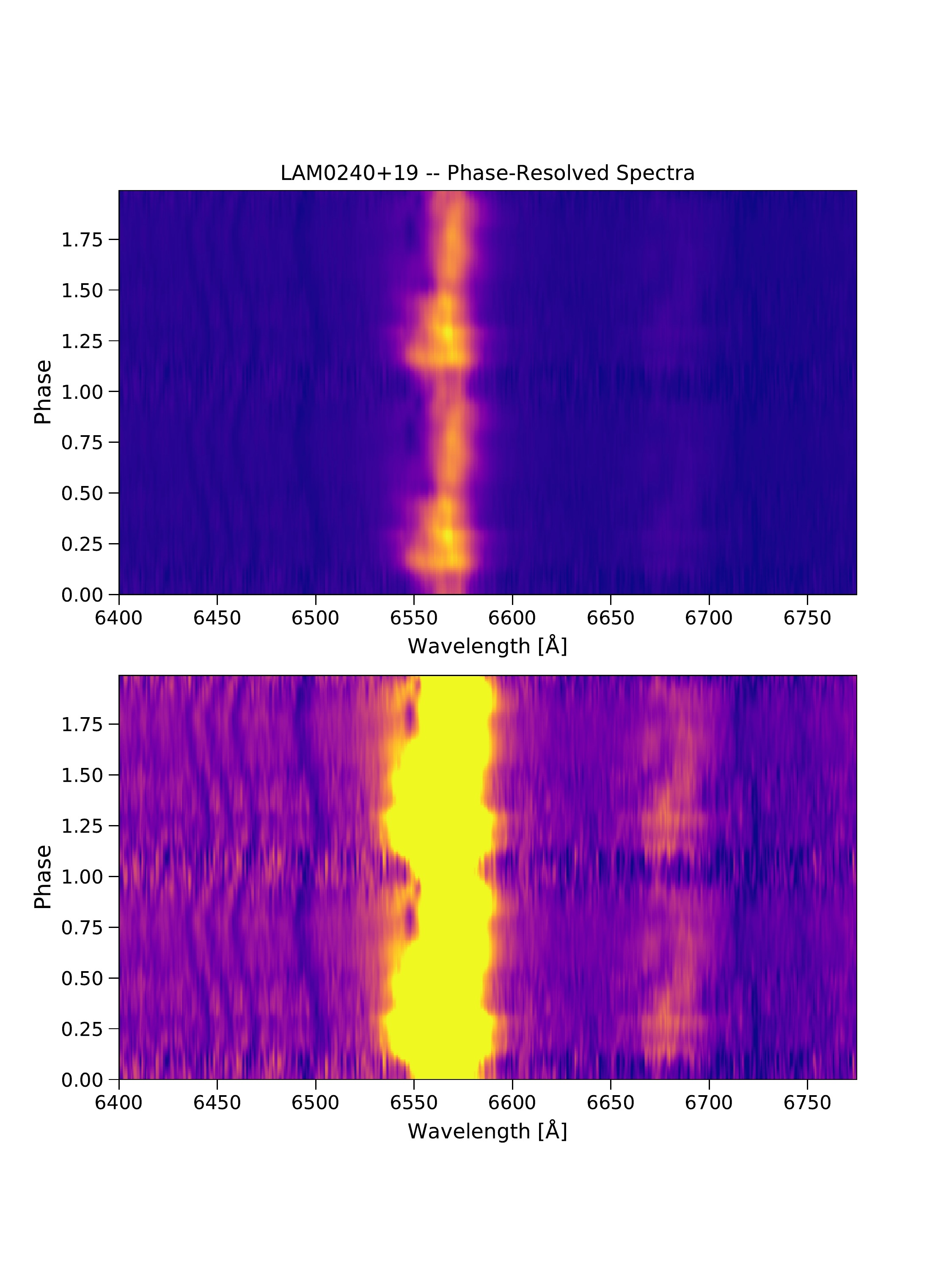}
\caption{Spectra of LAM 0240+19 near H$\alpha$, as a function of phase.
In the upper panel, the colormap is set to show details of the H$\alpha$ line,
and in the lower panel it is set to show the absorption features and the 
\ion{He}{1} $\lambda 6678$ emission.  The H$\alpha$ profile shows
irregular variations because all spectra are included, including 
those taken during flares.
}
\label{fig:lam0240+19trail}
\end{figure}

Fig.~\ref{fig:lam0240+19trail} shows a synthetic 
``trailed'' spectrogram.  Each line in the image
represents the spectrum at a phase in the orbit,
and is formed from an average of spectra taken 
near the fiducial phase, weighted by a 
Gaussian in phase.  The spectra were continuum-
divided (rectified) before averaging. 
Horizontally-extended noisy patches are 
poorly-covered intervals of orbital phase.
The absorption lines of the late-type star show
an obvious velocity modulation, but the emission
lines behave in a complicated fashion that does 
not appear repeatable with phase.  

\begin{figure}
\includegraphics[width=6.3 truein,trim = 1cm 1.3cm 1cm 2cm,clip]{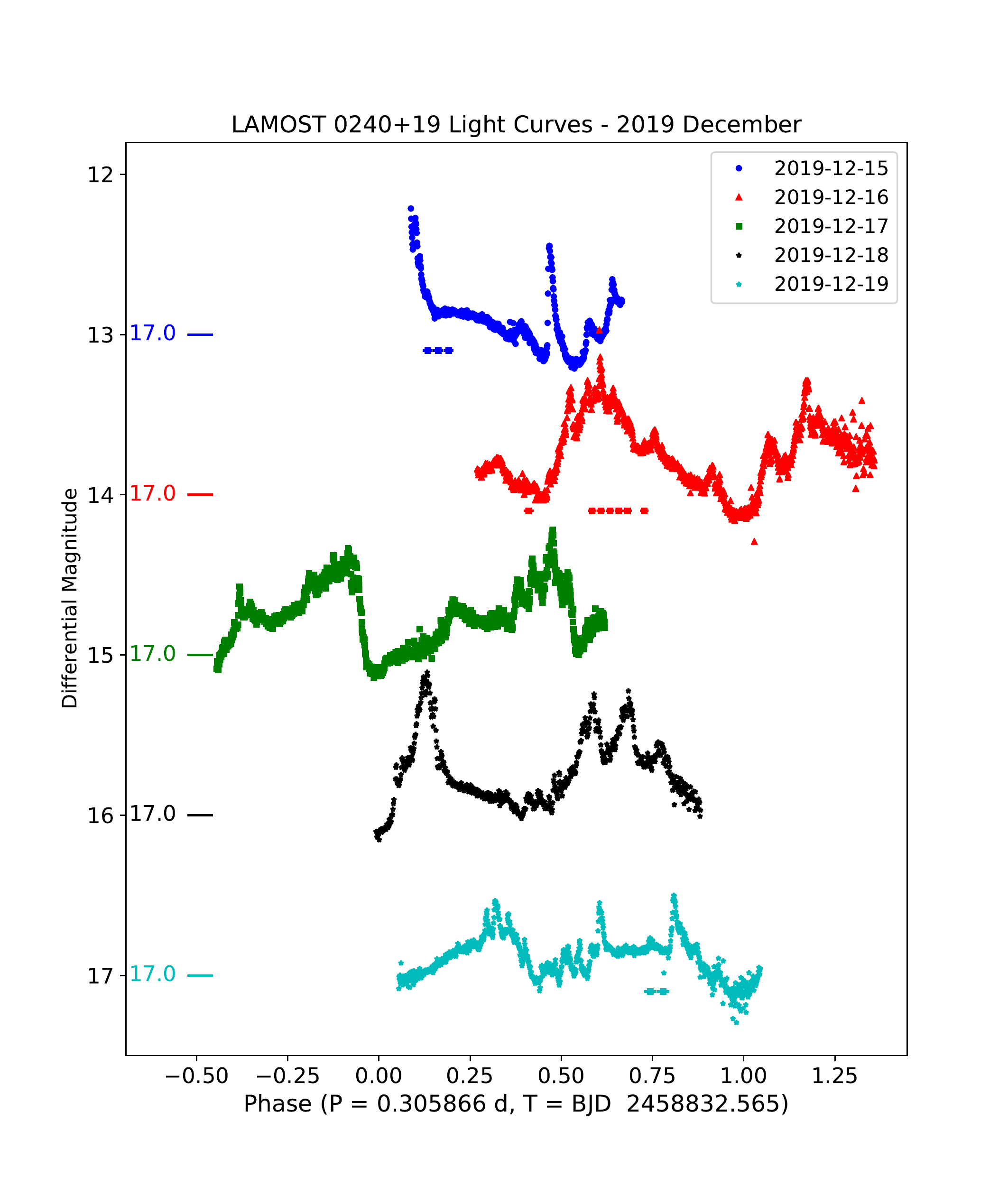}
\caption{  
Light curves of LAMOST 0240+19 from five successive nights in 
2019 December, plotted against orbital phase.  The zero points
for different nights are offset by 1.0 mag as indicated by the 
color-coded tick marks.  The white-light differential magnitudes
have been adjusted to rough $V$ magnitudes using the comparison
star, near $\alpha_{\rm ICRS}$ = 2:40:52.330, 
$\delta_{\rm ICRS}$ = +19:53:00.74,
for which the APASS catalog \citep{apass} gives $V = 15.91$.
Exposures were 20 s, with a cycle time of 23.3 s, except for
2019-12-18 UT, for which the corresponding figures 
were 30 s and 33 s.  Also shown are the times (orbital phases)
of spectroscopic exposures taken simultaneously; these 
are plotted as color-coded squares at the 17.1-magnitude level for 
each night.  The short horizontal error bars indicate the 
duration of each exposure.
}
\label{fig:lam0240+19lightcurves}
\end{figure}

Fig.~\ref{fig:lam0240+19lightcurves} shows time-series 
photometry from five
successive nights in 2019 December.
They show intermittent rapid flaring by almost 
1 magnitude, on timescales as short as $\sim 1$ min in the 
first light curve, and also flickering on longer time scales.
There is no obvious correlation with orbital phase in this
short data set, though the CRTS data suggest they
are more common near quadrature than near conjunction.  
The behavior is reminiscent of magnetic CVs, but nearly 
all magnetic CVs show strong \ion{He}{2} emission, 
which is absent here. 

\begin{figure}
\includegraphics[width=6.5 truein,trim = 0cm 1cm 1cm 2cm,clip]{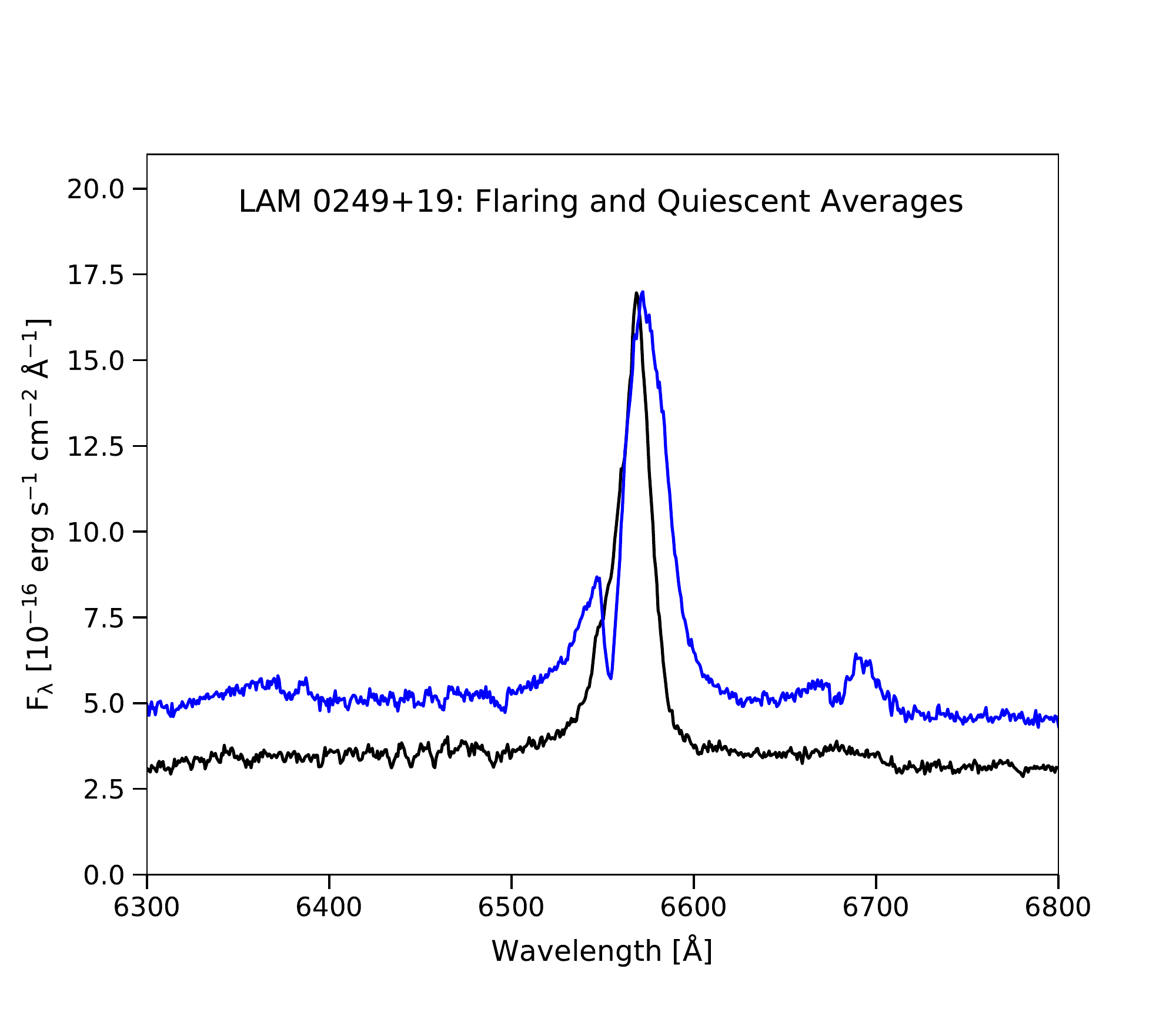}
\caption{  
Averages of spectra taken during and outside of flaring
intervals in the region of H$\alpha$.  As these were
taken, the comparison star used for the photometry
appeared reasonably steady, and the setup exposures
indicated similar seeing throughout, so the 
flux difference between the flaring (upper, blue trace) and
non-flaring (lower, black trace) is probably real.
}
\label{fig:lam0240+19flarespec}
\end{figure}

On three of the five nights for which we have photometry,
we obtained some spectra that were simultaneous.  The
time intervals covered are also shown in 
Fig.~\ref{fig:lam0240+19lightcurves}.  
Fig.~\ref{fig:lam0240+19flarespec} shows averages
of five spectra taken 2019 Dec.~16, when the 
simultaneous photometry showed the source flaring,
and four spectra taken Dec.~15 and 19, when 
flares were not occurring.
During the flaring interval, H$\alpha$ 
emission line grew stronger and broader,
and developed a blueshifted absorption feature 
at $\sim -420$ km s$^{-1}$.  H$\beta$ (not shown) 
behaved similarly.  The inconsistent emission-line
velocities noted earlier, and the chaotic appearance
of the emission lines in Fig.~\ref{fig:lam0240+19trail}, 
appear to be associated with flaring.

Unlike nearly all newly-discovered CVs, 
LAMOST 0240+19 does not naturally fall into one of the
well-populated CV subclasses.  It is evidently not a dwarf nova, nor
is it a polar (AM Her star) or intermediate polar (DQ Her
star), nor does it resemble the UX UMa or SW Sex-type novalikes.  
However, with the limited information we have available, it
is an excellent match for an object that has until 
now been unique -- AE Aquarii. In what follows we summarize
what is known about AE Aqr, and point out the apparent
family resemblance.

AE Aqr was among the first CVs recognized as a binary star
\citep{joy43}; it is bright, with a
mean $G = 10.95$ in the Gaia DR2.  Its brightness 
is largely caused
by its proximity (91 pc), which puts it in the $< 150$ pc 
volume-limited sample assembled by \citet{pala20}.
It is evidently a magnetic CV, but with unusual 
features. 

Like LAMOST 0240+19, AE Aqr is a longer-period CV
with a prominent secondary star contribution 
already noted by \citet{joy43}.  
Its orbital period, 9.98 hr, is not dissimilar to
the 7.34-hr period of LAMOST 0240+19.  As 
expected at this somewhat longer orbital period, 
its secondary star (K5; \citealt{robinson91})
is somewhat warmer than the M1.5 secondary in 
LAMOST 0240+19.  Also like LAMOST 0240+19, the spectrum 
of AE Aqr does not show appreciable \ion{He}{2} emission, and the 
\ion{He}{1} lines are also not strong for a 
CV \citep{williams83,echevarria89}.
While the emission lines in AE Aqr can vary rather
smoothly in velocity \citep{robinson91}, they often show dramatic 
and irregular variation \citep{chincarini81, reinsch94, 
welsh98}.

The most striking similarity between AE Aqr and 
LAMOST 0240+19 is that their light curves
both show distinctive, irregular
flaring on time scales of minutes.  Nearly all CVs 
flicker, but flares of this kind are unusual except 
among magnetic CVs in high states.  The literature 
offers many examples of flaring light curves of 
AE Aqr (e.g. \citealt{chincarini81,
welsh98, watson06}).  The resemblance 
between these and Fig.~\ref{fig:lam0240+19lightcurves}
is striking.

At high time resolution, the light curve of AE Aqr
shows a coherent 33-s oscillation \citep{patterson79}, 
which is thought
to be the rotation period of a magnetized white dwarf.
This is the shortest known white dwarf rotation period.
The unique features of AE Aqr are explained by 
a {\it propeller effect} -- the magnetic field of the 
rapidly spinning white dwarf expels most of the 
mass transferred from the secondary, with only a 
rather small fraction ultimately accreting onto the 
white dwarf \citep{eracleous96,wynn97}. 

If high time resolution photometry were to yield
a short-period (tens of seconds) coherent 
pulsation in LAMOST 0240+19, its kinship
to AE Aqr would be conclusively proven.
Unfortunately, the cadence of our photometry
-- 23 sec in 2019 December and 30 sec for a
short segment in 2020 January -- is too slow
to show this.  Searches at very high frequencies
show only aliases of the ellipsoidal modulation.
When new observations become possible, a search
for pulsations in the relevant period range
could be very rewarding.  LAMOST 0240+19 is much
fainter than AE Aqr, and the pulsations in AE Aqr
are not always obvious, but the observational
task is eased by the coherence of white
dwarf rotation over long time spans.

\subsection{LAMOST J033940.98+414805.7}
\label{subsec:lam0339}

\begin{figure}
\includegraphics[width=7.5 truein, trim=0.5cm 4cm 0 5cm,clip]{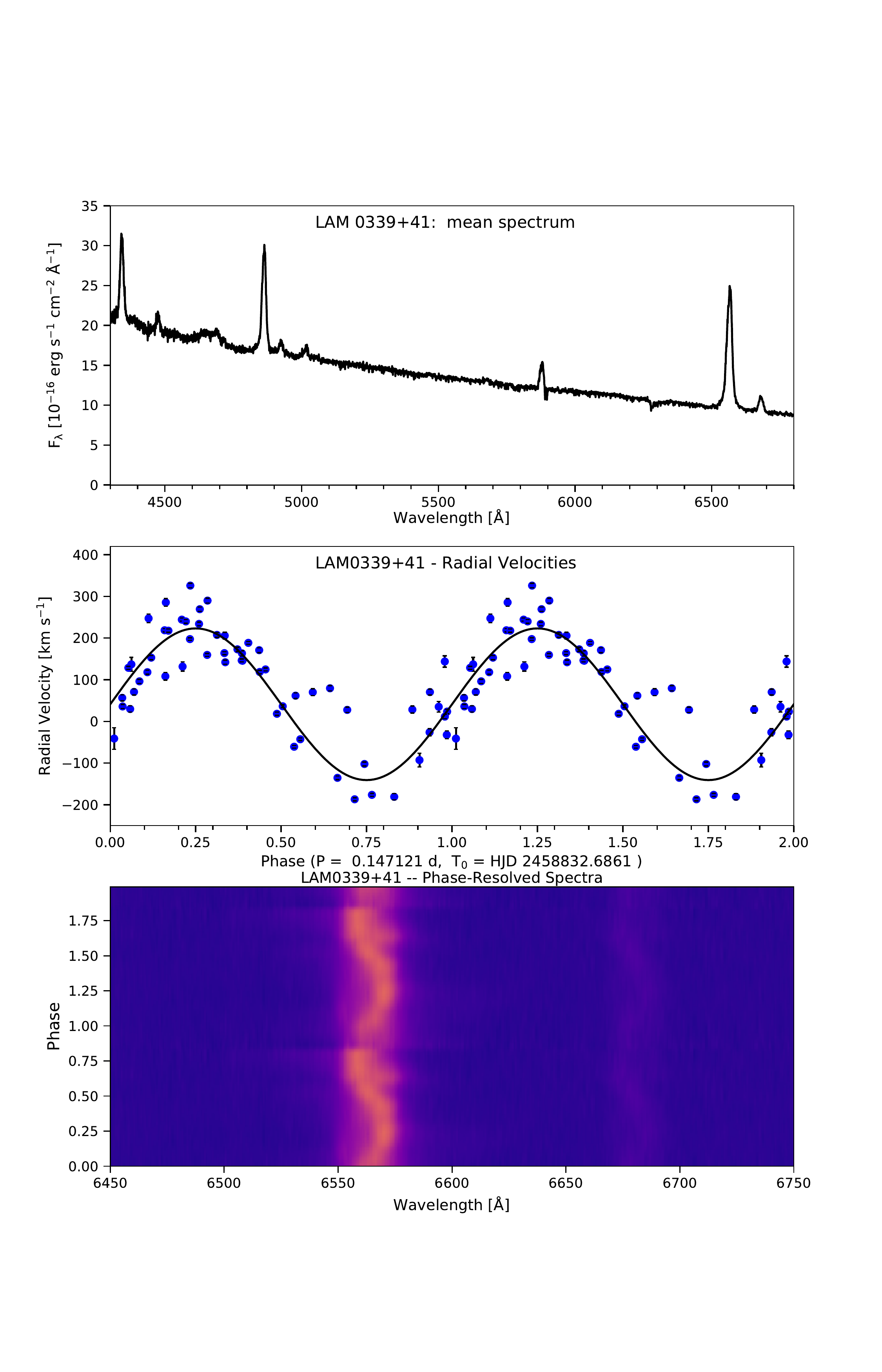}
\caption{  
Upper panel: The mean spectrum of LAMOST 0339+41. 
Middle panel:  Radial velocities of the narrower 
component of the H$\alpha$ emission line, folded on the 
ephemeris shown.  Lower panel: A portion of the 
rectified, phase-averaged spectrogram.}
\label{fig:lam0339montage}
\end{figure}

The mean spectrum of this object (Fig.~\ref{fig:lam0339montage},
top panel) is typical for a novalike CV, with a strong, blue
continuum and emission lines of hydrogen and \ion{He}{1}.
Weak emission is visible at \ion{He}{2} $\lambda 4686$ and the 
\ion{C}{3}-\ion{N}{3} blend at $\lambda 4640$.  Its position
is not covered by the CRTS.

We measured radial velocities of the H$\alpha$ emission using
for the convolution function 
the derivative of a Gaussian,
optimized for a line with a 900 km s$^{-1}$ FWHM.  
The middle panel of Fig.~\ref{fig:lam0339montage} shows the 
velocities folded on a period of 3.543(5) hr, which
is determined without ambiguity.  Many CVs in this period range are 
SW Sex stars \citep{r-g07,thor_swsex91}, and the mean spectrum
does appear similar to stars of this class; however,
the phase-averaged spectrogram shown in the
lower panel of Fig.~\ref{fig:lam0339montage} does not 
show the phase-dependent absorption often present in
SW Sex stars.  The appearance of phase-dependent absorption
probably requires an orbital inclination not too far
from edge-on.  The H$\alpha$ line in the phased
spectrogram appears to have two components, one
of them fairly narrow and the other broader.  The
narrower component appears to be the one measured
by the convolution function; the broad component shows
little or no motion on the orbital period.  The narrower
component may arise from the bright spot formed 
where the mass-transfer stream strikes the accretion
disk; if so, most of its velocity variation would be
from the rotation of the disk rather than the orbital
motion of the white dwarf.

In summary, this appears to be a fairly typical, but
previously unrecognized, novalike variable.

\subsection{LAMOST J035913.61+405035.0}
\label{subsec:lam0359}

\begin{figure}
\includegraphics[width=7.0 truein,trim = 0cm 1cm 0cm 2cm,clip]{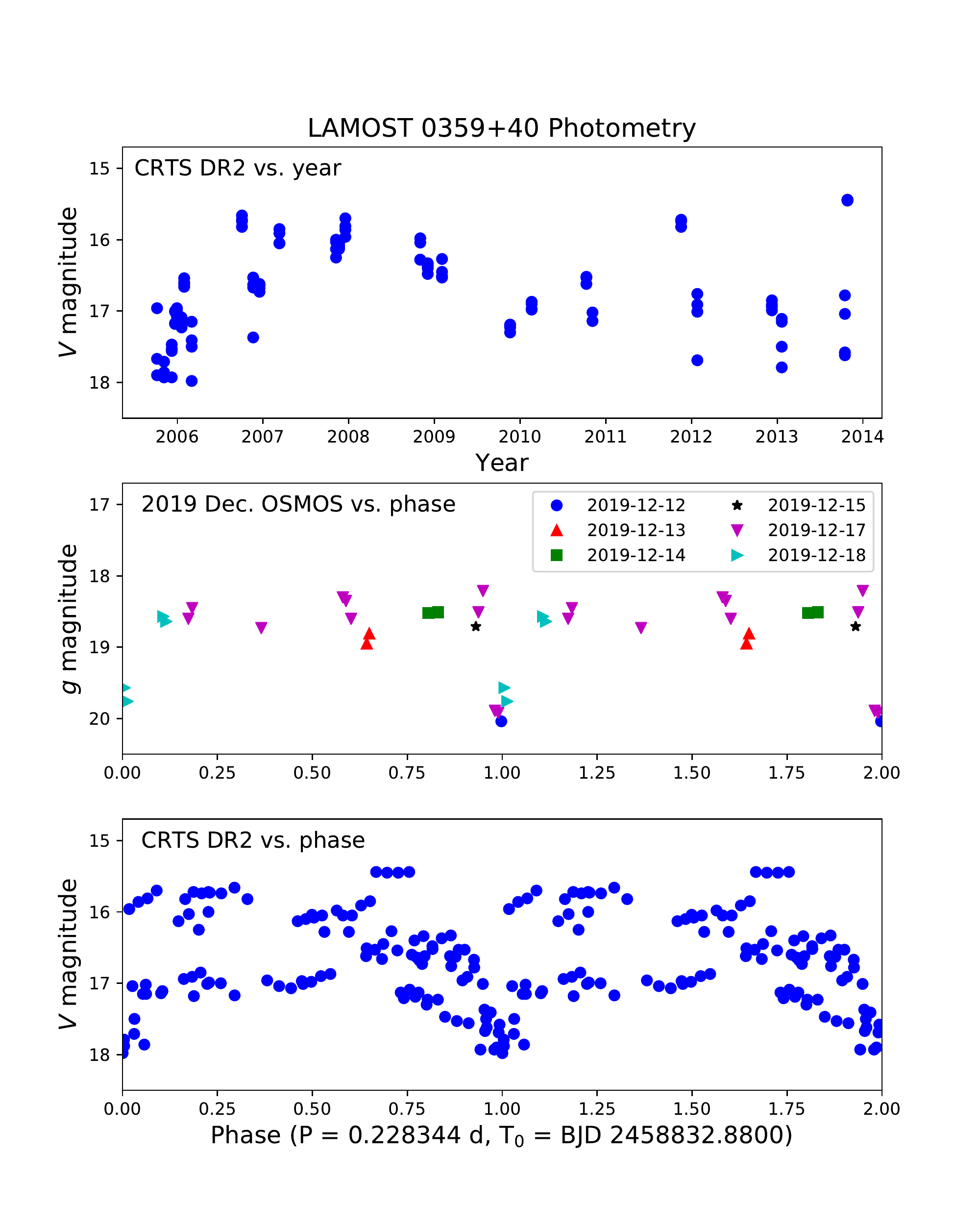}
\caption{  
Top panel: Archival photometry of LAM 0359+40 from the CRTS DR2. 
Middle panel: Magnitudes from target acquisition images taken 2019 Dec., 
folded on the provisional eclipse ephemeris.  Lower panel: The same
data as the top panel folded on the provisional ephemeris.
}
\label{fig:lam0359phot}
\end{figure}

The top panel of Fig.~\ref{fig:lam0359phot} shows 
the CRTS DR2 light curve 
of this object, which appears to have spent most of 
2007-2009 in a relatively bright state around 16th
magnitude, and then faded to a more 
typical state near 17th, with occasional 
outbursts to the brighter state.  The light 
curve is consistent with a dwarf nova; the
persistent high state is characteristic of the Z Cam
subclass.  The total range of variation, $\sim
2.5$ mag, is rather low for a dwarf nova.

\begin{figure}
\includegraphics[width=7.0 truein, trim = 0cm 1.5cm 0cm 2cm,clip]{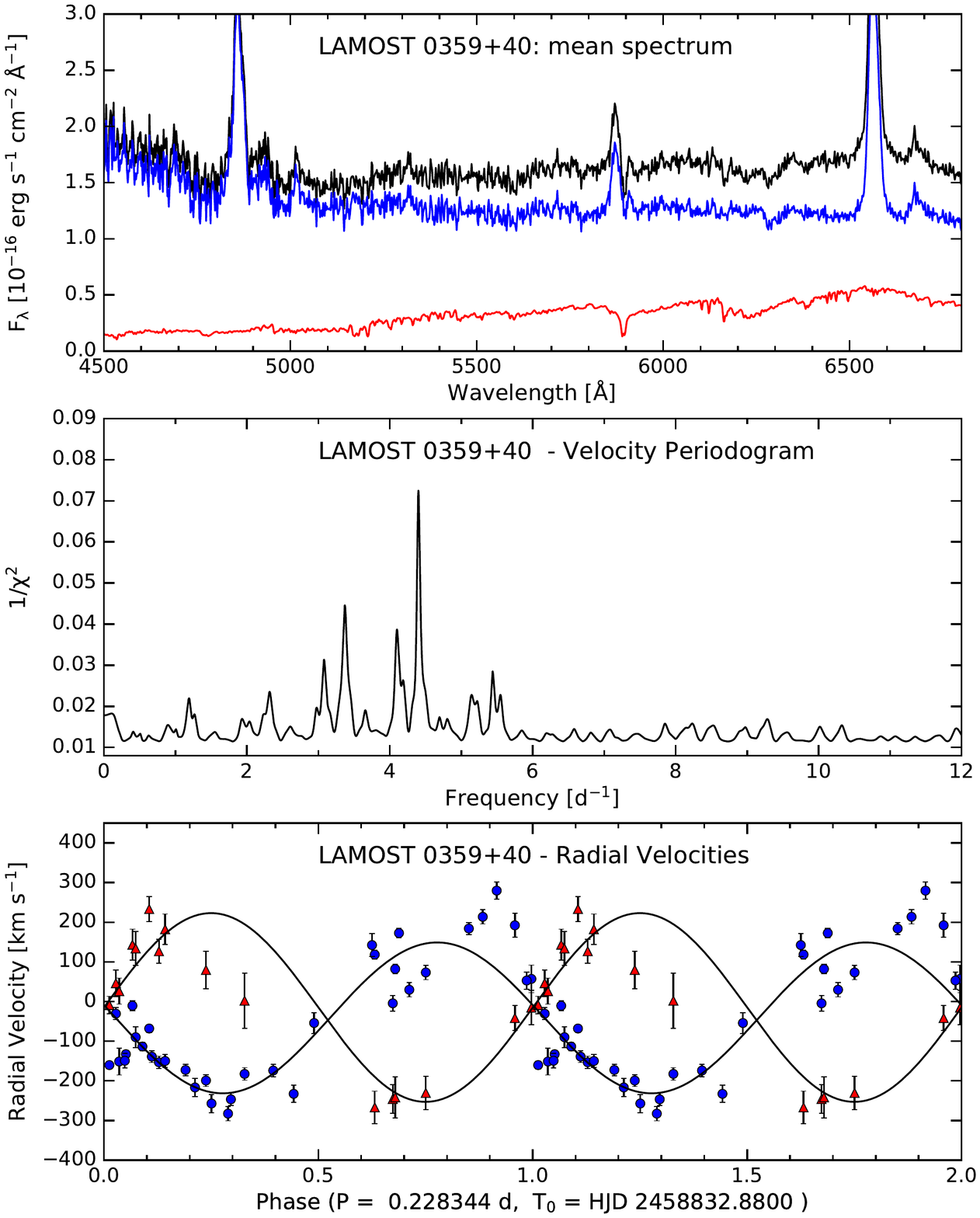}
\caption{  
Top panel: The top trace (black) shows the mean spectrum from 2019 December.
The lower trace (red) is a spectrum of the M1.0 star Gliese 514, scaled to 
$V = 20.3$ mag.  The middle trace (red) is the difference between these two.
Middle panel: Periodogram of the H$\alpha$ emission velocities. Lower panel:
Radial velocities of the emission (blue) and absorption (red),  
folded on the provisional ephemeris (see text).  The best-fitting
sinusoids are overplotted; in the fit shown for the absorption, the phase
was fixed to that of the eclipse ephemeris.
}
\label{fig:lam0359montage}
\end{figure}

Our mean spectrum (Fig.~\ref{fig:lam0359montage}, top panel) appears
typical for a dwarf nova near minimum light.  An early M star contributes
to the continuum, as can be seen from the absorption bands longward of
6000 \AA .  We estimate the spectral type as M1 $\pm$ 1 subclass, and
that the secondary star's contribution is equivalent to $V = 20.3$,
with an uncertainty of $\sim 0.5$ mag.  A period search of the  
H$\alpha$ emission-line velocities (Fig.~\ref{fig:lam0359montage})
shows an unambiguous period of 5.494(17) hr.  We attempted to measure
cross-correlation velocities of the M dwarf, but due to the faintness
of its contribution and sometimes poor signal-to-noise, only some of
our spectra gave usable results, and those were rather imprecise. 
The lower panel of Fig.~\ref{fig:lam0359montage} shows both the emission 
and cross-correlation velocities folded on the provisional eclipse ephemeris
(discussed below).

The target acquisition images mostly showed the source near $g = 18.5$ in
2019 December, but four images showed it distinctly fainter at 
$g > 19.5$ (Fig.~\ref{fig:lam0359phot}, middle panel).  Faint
states occurred on three different nights, and always occurred
near the same orbital phase.  Furthermore, the phase of these
faint states corresponds approximately to blue-to-red
crossing of the M-dwarf velocity, which is the inferior conjunction
of the red dwarf.  The faint state is evidently an eclipse.

We attempted to refine the ephemeris by folding the CRTS data 
on a range of finely-spaced periods near the period found
from the H$\alpha$ velocities.  
We found several refined periods in which the faint CRTS points
cluster in phase; the best-fitting of these gives a provisional
eclipse ephemeris 
\begin{equation}
\hbox{BJD eclipse} = 2458832.880 + 0.228344 E,\ \ \hbox{[provisional]},
\label{eqn:lam0359provisional}
\end{equation}
where $E$ is an integer cycle count and the quantities are uncertain
by one or two in the last digits shown.  The lower panel of 
Fig.~\ref{fig:lam0359phot} shows the CRTS data folded on the
provisional ephemeris.  

The provisional ephemeris should be easy to test
and refine with precise eclipse timings from time-series
photometry resolving the ingress and egress.
If more accurate radial velocities of the secondary star 
can also be obtained, a good white-dwarf mass estimate 
should be within reach.

\subsection{LAMOST J090150.09+375444.3}
\label{subsec:lam0901}

\begin{figure}
\includegraphics[width=7.0 truein, trim=0cm 1.5cm 0cm 2cm, clip]{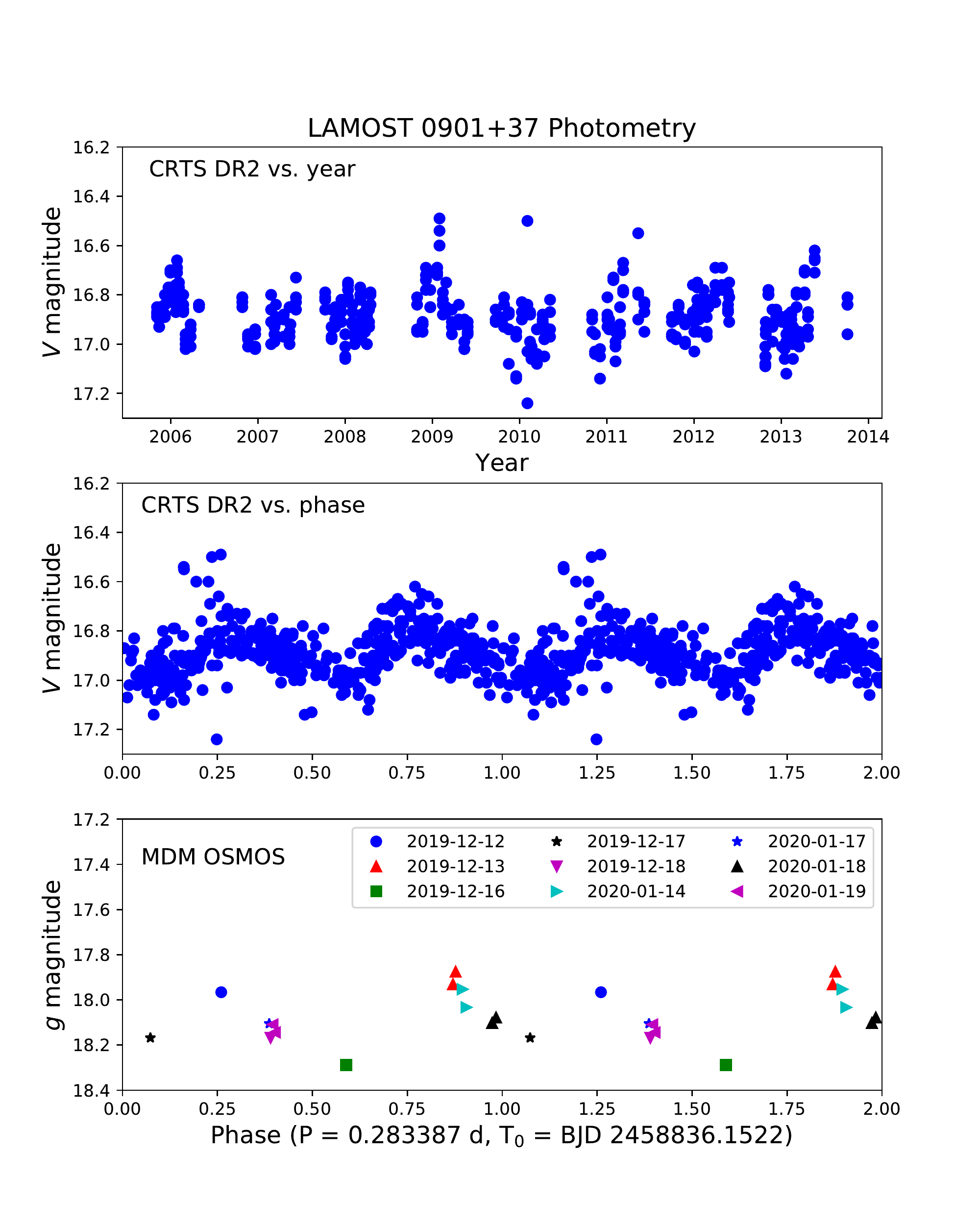}
\caption{Top panel: Archival CRTS magnitudes of LAMOST 0901+37 
plotted as a function of time.  Middle panel: The same data
plotted as a function of phase.  The period is from \citet{drake14},
and the epoch chosen to match the blue-to-red crossing of the 
absorption line radial velocities.  Lower panel: $g$ magnitudes
from the OSMOS target acquisition images, plotted against phase.
}
\label{fig:lam0901phot}
\end{figure}

This object's LAMOST spectrum \citep{hou20} shows strong, rather narrow 
Balmer and HeI emission lines, and \ion{He}{2} $\lambda$4686
about 2/3 of the strength of H$\beta$, which is usually
a sign of magnetic activity.  A sizable late-type star
contribution is also visible in the continuum, with 
the strong blend and continuum break near 5180 \AA\ 
characteristic of late K stars, and relatively weak
molecular absorption bands typical of early M stars.
It appears in the list of periodic variables found 
in CRTS; \citet{drake14} list $P = 0.2833870$ d.
The top panel of Fig.~\ref{fig:lam0901phot} shows
the CRTS data; the bottom panel shows the same
data folded on the period, showing the double-humped
signature of ellipsoidal modulation, with a full 
amplitude of a bit over 0.2 magnitude.  No substantial 
outbursts are observed.  The variation in the 
$g$ magnitudes from the target-acquisition images 
(lower panel) appears consistent with the ellipsoidal
modulation in the CRTS data.

\begin{figure}
\includegraphics[width=7.0 truein,trim = 0cm 1cm 0cm 1.5cm, clip]{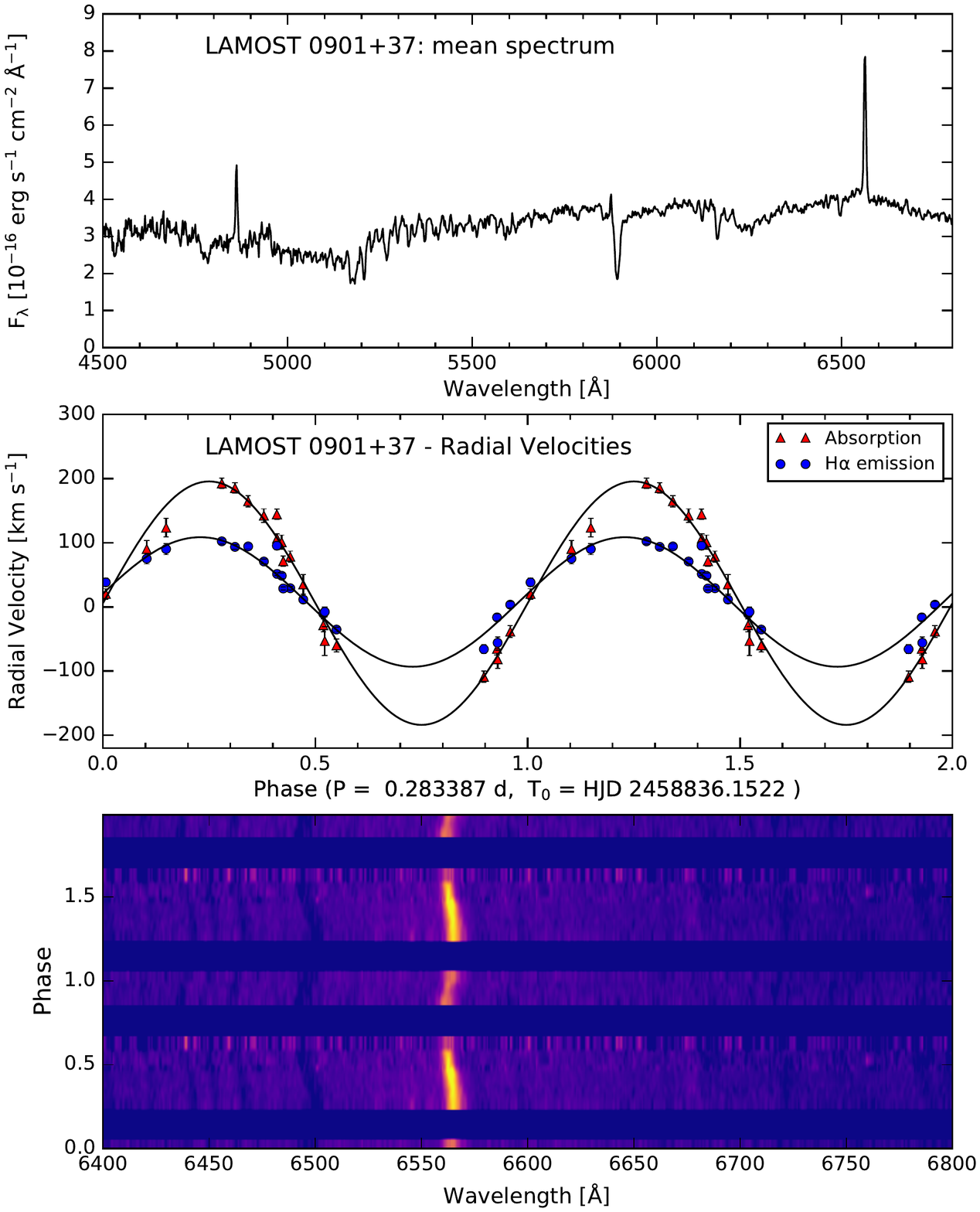}
\caption{Top panel: Mean MDM spectrum of LAMOST 0901+37, from 
2019 December and January.  Middle panel: Radial velocities of the 
H$\alpha$ emission (blue) and late-type absorption features (red)
plotted as a function of phase.  Lower panel: Phase-averaged
spectra in the vicinity of $H\alpha$.  The blank horizontal
spaces correspond to phases that were not covered.}
\label{fig:lam0901montage}
\end{figure}

Our mean spectrum (Top panel of Fig.~\ref{fig:lam0901montage})
is significantly different from the LAMOST spectrum,
in that \ion{He}{2} $\lambda$4686 is not detected, and the 
\ion{He}{1} lines are also less prominent.  The late-type
contribution is still present.  We tried our standard procedure
of scaling and subtracting late-type spectra from this, but
this did not give a satisfactory spectral decomposition.  
The strong absorption in the 
$\sim 5150$ \AA\ region could only be subtracted with a strong
late-K contribution, but this did not do a good job on the 
TiO bands toward the red.  Early M-type spectra fit the
TiO bands well, but did not match the shorter-wavelength
features.

Both the H$\alpha$ emission and the absorption velocities
varied smoothly with orbital phase, and both data sets
independently yielded periods consistent with the more 
precise \citet{drake14} period.  The blue-to-red
crossing of the absorption velocities corresponds to
inferior conjunction of the late-type star, so we adopt
this as phase zero for the ephemeris 
\begin{equation}
\hbox{BJD inferior conjuntion} = 2458836.152(2) +  0.2833870 E.
\end{equation}

The lower two panels of Fig.~\ref{fig:lam0901montage} show
that the H$\alpha$ emission velocities move
{\it in phase} with the absorption, with a somewhat smaller
velocity amplitude.  It appears to arise from the side of
the secondary star that faces the primary.  Emission 
lines from the secondary are seen in some CVs, but seldom 
dominate the emission.  

Several scenarios could explain this unusual behavior.  
In all the scenarios, the rate of accretion onto the primary must have
been low enough during our observations so that we do not 
see an accretion disk or column.  

\begin{figure}
\includegraphics[width=6.5 truein]{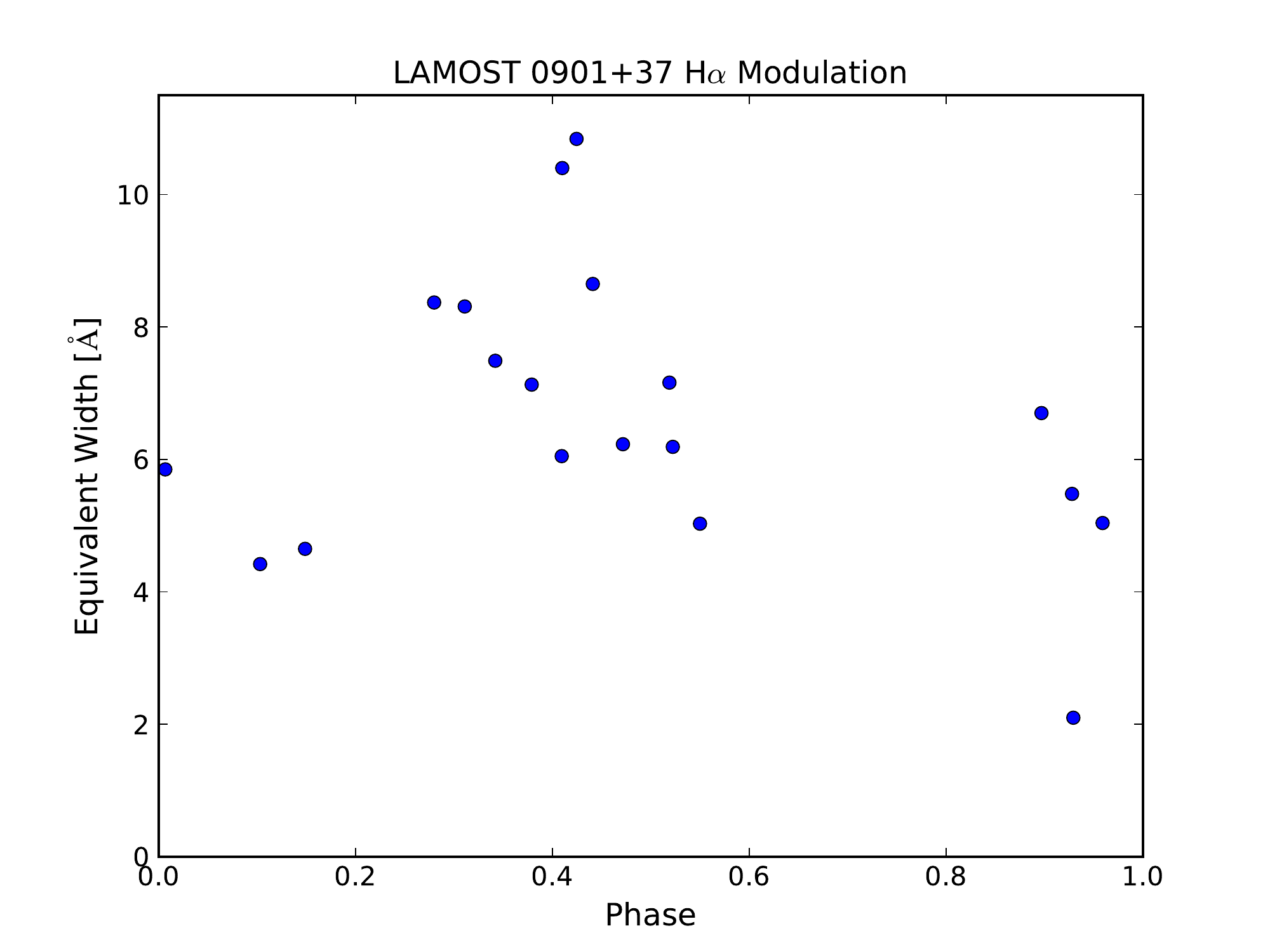}
\caption{Equivalent width of the H$\alpha$ emission in
LAMOST 0901+37 as a function of orbital phase.}
\label{fig:lam0901ewfig}
\end{figure}

One explanation is that the primary star emits enough
ionizing radiation to excite fluorescent emission on 
the facing side of the secondary.  However, if this mechanism
dominates the emission, and the flux of ionizing radiation 
is steady and isotropic, the changing aspect of the illuminated
face should cause the emission line strength to vary 
smoothly and symmetrically with orbital phase;
the detached WD-dM binary FS Cet (Feige 24; \citealt{tcmb78}) 
remains an excellent example.  Fig.~\ref{fig:lam0901ewfig} 
shows the H$\alpha$ emission line strength as a function of orbital phase.
The illuminated-face hypothesis predicts maximum strength
at phase 0.5 (red-to-blue crossing for the secondary 
stars's velocities), and while there is a tendency for 
the line to be stronger around phase 0.5, the variation is
not as smooth and symmetric as expected. 

Alternatively, the magnetic activity of the secondary may
be strong.  CV secondary stars are
forced to rotate rapidly because they are 
tidally locked to the orbit.  The rapid rotation is thought to enhance
their magnetic activity, which in turn should produce
chromospheric emission.  In most CV spectra the accretion
makes this emission inconspicuous, but it could emerge when 
the accretion turns off.  The chromospheric hypothesis
does not naturally explain why the 
velocity amplitude of the emission is smaller than
that of  the absorption; explaining this requires 
an {\it ad hoc} assumption that the emission arises 
mostly on the side facing the primary.

The velocity modulation does prove that LAMOST 0901+37 
is a close binary system, but our spectra alone do not 
demonstrate that it is a CV.  However, the strong 
\ion{He}{2} $\lambda$4686 in the \citet{hou20} spectrum 
does show conclusively that the system goes into more
active states, and a strong $\lambda$4686 line often indicates
a magnetic CV.  These sometimes drop into 
low states in which the accretion practically stops.  The
low states can last for years (see, e.g., EF Eri; \citealt{szkody10_eferi}).
Another class of stars in which mass transfer essentially shuts
off are the VY Sculptoris stars, a class of novalikes that strongly 
overlaps SW Sex stars.   VY Scl stars can spend long intervals
in low states, but typically have periods in the 
3-4 hr range \citep{rodriguez20}, much shorter than here.

LAMOST 0901+37 may be similar to V405 Peg \citep{thor_rbs1955_09,schwope14}, 
which has a slightly shorter period but 
also has a varying emission spectrum.  In V405 Peg, 
the H$\alpha$ line in the strong-line state varies 
in velocity roughly in antiphase to the absorption-line 
spectrum, while in the weak-line state H$\alpha$ is narrower 
and its motion small and ill-defined. A significant
difference between the two systems is that V405 Peg 
is a fairly strong X-ray source (Rosat Bright Source
1955), while none of the X-ray source catalogs on Vizier list 
any sources near ($< 2$ arcmin) LAMOST 0901+37.
At 525 pc, LAMOST 0901+37 is almost exactly 3 times as
distant as V405 Peg (173 pc in Gaia DR2
\footnote{\citet{thor_rbs1955_09} estimated the distance of 
V405 Peg to be $149(+26,-20)$ pc based mostly on ground-based 
parallax measurements.}, so if it were the same 
X-ray luminosity as V405 Peg it would be 
9 times fainter.  The ROSAT all sky bright source catalog
\citep{voges99} lists V405 Peg at 0.21 counts s$^{-1}$ and 
includes sources brighter than 0.05 counts s$^{-1}$, so it seems
possible that LAMOST 0901+37 has been missed because of distance.
It could also have a low duty cycle.

\subsection{LAMOST J204305.95+341340.6}
\label{subsec:lam2043}

\citet{margoni84} found the variability of this object 
in a Schmidt survey of a low-Galactic-latitude field.  They
list its magnitude range as $15.0 < V < 15.4$ and $16.0 < B < 16.9$, without
further detail.  \citet{skiff97} published improved coordinates.  
Its celestial location is not covered by the CRTS.  

\begin{figure}
\includegraphics[width=7.0 truein, trim = 0cm 4cm 0cm 4cm,clip]{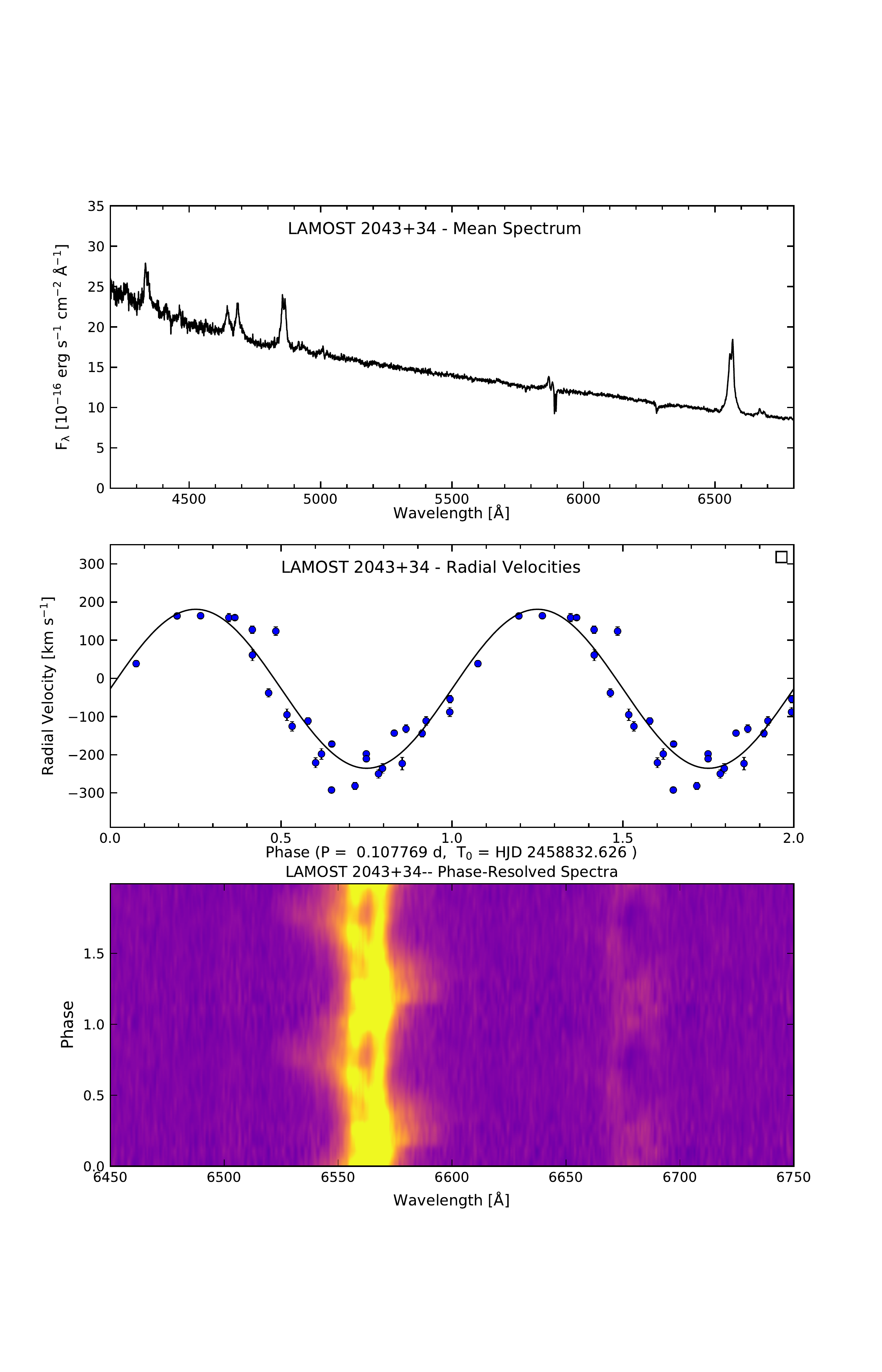}
\caption{  
Top panel: Mean spectrum of LAMOST 2043+34 from 2019 December.
Middle panel: Radial velocities of H$\alpha$ folded on the best-fitting
period of 2.586 hr; an alias period at 2.899 hr fits nearly as well.
Lower panel: Phase-averaged spectra in vicinity of H$\alpha$ and
\ion{He}{1} $\lambda$6678. 
}
\label{fig:lam2043montage}
\end{figure}

The LAMOST spectrum
\citep{hou20} has a strong blue continuum with broad Balmer 
emission lines and relatively weak \ion{He}{1} emission; 
however, \ion{He}{2} $\lambda 4686$ and the \ion{C}{3}-\ion{N}{3}
blend around $\lambda 4340$ are both strong, around half the
strength of H$\beta$.  
The mean MDM spectrum (top panel of Fig.~\ref{fig:lam2043montage}) is
similar to the LAMOST spectrum.  The target acquisition
images have $15.39 < g < 15.71$, indicating that the photometric
state was similar to that in which \citet{margoni84} found it.

A search of the H$\alpha$ velocities  
yields a best-fit period of 2.586(3) hr, but because the 
object was only observable for a few hours per night,
a 2.899(3) hr period also fits well; these
two periods differ in frequency by almost 
exactly 1 cycle d$^{-1}$. 
The middle panel of Fig.~\ref{fig:lam2043montage} shows
the velocities folded on the 2.586(3) hr period.
While the daily cycle count remains undetermined, 
the system's period evidently lies in
the so-called 2-3 hour ``gap" in the CV 
period distribution.

The lower panel of Fig.~\ref{fig:lam2043montage} shows
the phase averaged spectra near H$\alpha$.  The central
part of the line profile has two distinct peaks separated
by $\sim 550$ km s$^{-1}$.  The peaks vary relatively little in velocity;
by marking the peaks on an expanded display of the data in 
Fig.~\ref{fig:lam2043montage}, we estimate $K \sim 60$ km s$^{-1}$
for the peaks' motion.
The most striking feature of the 
line profile is a much wider wing that swings from one side of the 
line to the other over the orbit.
At its maximum red and blue shifts, near phases 0.3 and 0.8 in the
figure, it extends $\sim 2000$ km s$^{-1}$ from rest.
The pattern strikingly 
resembles the line profile variation in 
V795 Her, for which \citet{casares96} present time-series
spectroscopy and a thorough analysis. 
V795 Her is a novalike variable with $P_{\rm orb} = 2.597$ hr,
very close to the 2.586 hr best-fit period
for LAMOST 2043+34.
The similarity extends to the two systems' luminosities as well;
Gaia DR2 list 589(13) pc and G = 13.08 for V795 Her, and \citet{green19} 
give $E(g-r)$ = 0.04(2) mag.  This results in $M_G = 4.1$ for V795 Her, 
while the same calculation yields $M_G = 4.3$ for LAMOST 2043+34 
(Table~\ref{tab:targets}).  
LAMOST 2043+34 and V795 Her appear to be virtually identical.

\section{Discussion}
\label{sec:discussion}

These results underscore how easy it can be for interesting
objects to go unnoticed, if the usual search criteria do not
include them.
While three of the five stars studied here had previously been 
noted as variable, none have a large range of variation, and none
had been recognized as a CV.  Even so, all of the stars prove
to be interesting and for several, further work should yield
significant improvements.  Our most surprising result 
is that LAMOST 0240+19 appears to be the long-lost twin of AE Aqr;
this cries out for fast photometry sufficient to resolve 
pulsations with periods down to a few seconds.
LAMOST 0359+40 has a detectable
secondary star and an eclipse, and may be favorable for a 
white-dwarf mass determination; this will require proper
time-series photometry to resolve the eclipses and sharpen the
ephemeris, and better velocities of the secondary.
LAMOST 0901+37 appears to have dipped into a low state in which 
accretion has dropped to negligible levels; it should be 
watched for a return to the high state seen in its
discovery spectrum.  LAMOST 2043+34 is in the period gap, 
and appears identical to V795 Her.  LAMOST 0339+41
appears to be a low-inclination SW Sex star; these are often
also VY Scl stars, so it may dive into low states in the future.


\acknowledgments

The CSS survey, which is funded by the National Aeronautics and Space
Administration under Grant No. NNG05GF22G issued through the Science
Mission Directorate Near-Earth Objects Observations Program.  The CRTS
survey is supported by the U.S.~National Science Foundation under
grants AST-0909182 and AST-1313422. In addition, this research made 
use of the AAVSO Photometric All-Sky Survey (APASS), funded by the 
Robert Martin Ayers Sciences Fund and NSF AST-1412587.

We thank Dartmouth graduate students Kathryn Weil and especially
Aylin Garcia-Soto for helping with the 2019 December observations,
as well as Dartmouth undergraduates 
Catherine Slaughter, Chase Alverado-Anderson,
and Piper Stacey for their company and assistance.  Heartfelt 
thanks go to Claudia and Jay Weed for underwriting 
undergraduate observatory travel. 

Finally, we would like to thank the Tohono O'Odham Nation
for leasing their mountaintop, so that we may explore the 
universe that surrounds us all.


\bibliographystyle{yahapj}
\bibliography{ref}

\end{document}